\documentclass[prl,floatfix,superscriptaddress,showpacs,twocolumn]{revtex4}
\usepackage{graphics}
\usepackage{graphicx}
\usepackage{dcolumn} 
\usepackage{bm}
\usepackage{epsfig}
\usepackage{amsmath}
\usepackage{amsfonts}
\usepackage{amssymb}
\usepackage{times}
\newcommand{\eg}{{\sl e.g.}}

\newcommand{\sto}{SrTiO$_3$} 

\newcommand{\lao}{LaAlO$_3$}

\newcommand{\muB}{$\mu_{\rm B}$}

\begin{document}

\title{
Tuning the two-dimensional electron gas at the LaAlO$_3$/SrTiO$_3$(001) interface by metallic contacts
}

\author{Victor G. \surname{Ruiz L\'opez}}
\affiliation{Department of Earth and Environmental Sciences, Section Crystallography and Center of Nanoscience,
University of Munich, Theresienstr. 41, 80333 Munich, Germany}
\author{R\'emi Arras}
\affiliation{Department of Earth and Environmental Sciences, Section Crystallography and Center of Nanoscience,
University of Munich, Theresienstr. 41, 80333 Munich, Germany}
\author{Warren E. Pickett}
\affiliation{Department of Physics, University of California, Davis, One Shields Avenue, Davis, CA 95616, U.S.A.}
\author{Rossitza Pentcheva}
\email{rossitzap@lmu.de}
\affiliation{Department of Earth and Environmental Sciences, Section Crystallography and Center of Nanoscience,
University of Munich, Theresienstr. 41, 80333 Munich, Germany}

\date{\today}

\begin{abstract}
First principles calculations reveal that adding a metallic overlayer on LaAlO$_3$/SrTiO$_3$(001) eliminates the electric field within the polar LaAlO$_3$ film and thus suppresses the thickness-dependent insulator-to-metal transition observed in uncovered films. Independent of the LaAlO$_3$ thickness both the surface and the interface are metallic, with an enhanced interface carrier density relative to LaAlO$_3$/SrTiO$_3$(001) after the metallization transition. Moreover, a monolayer thick  metallic Ti-contact exhibits a finite magnetic moment and for a thin SrTiO$_3$-substrate induces a spin-polarized 2D electron gas at the \textit{n}-type interface due to confinement effects. A diagram of band alignment in $M$/LaAlO$_3$/SrTiO$_3$(001) and  Schottky barriers for $M$=Ti, Al, and Pt are provided. 
\end{abstract}

\maketitle

 The (001) interface between the band insulators \lao \ (LAO) and \sto \ (STO) provides remarkable examples of the novel functionalities that can arise at oxide interfaces, including  a quasi two-dimensional electron  gas (q2DEG)\cite{Ohtomo:Hwang2004}, superconductivity \cite{Reyren:Thiel:Mannhart2007}, magnetism~\cite{Brinkman:Huijben:etal2007} and even signatures of their coexistence~\cite{Dikin,Li}. Moreover, a thickness-dependent transition from insulating to conducting behavior was reported in thin \lao\ films on \sto(001) at $\sim4$ monolayers (ML) LAO~\cite{Thiel:Hammerl:Mannhart2006}. This  insulator-metal transition (MIT) can be controlled reversibly via an electric field, e.g. by an atomic force microscopy (AFM) tip~\cite{Cen:Thiel:Hammerl:etal2008,Bi:Levy:2010,Chen:2010}. Recently, it was shown that an additional STO capping layer can trigger the  MIT already at 2 ML of LAO, allowing stabilization of an electron-hole bilayer~\cite{Pentcheva:2010}. Density functional theory calculations (DFT)  have demonstrated the emergence of an internal electric field for thin polar LAO overlayers\cite{Ishibashi:Terakura2008,Pentcheva:Pickett2009,Pentcheva:PickettRev2010,Son:2009}. Screening of the electric field by a strongly compensating lattice polarization in the LAO film\cite{Pentcheva:Pickett2009,Willmott:2011} allows several layers of LAO to remain insulating before an electronic reconstruction takes place at around  4-5 monolayers of LAO. Recent AFM experiments provide evidence for such an internal field in terms of a polarity-dependent asymmetry of the signal~\cite{Xie:Hwang:2010}, but x--ray photoemission studies\cite{Segal:Reiner:etal2009,Sing:Claessen:2009,Chambers:Ramasse:2010} have not been able to detect shifts or broadening of core-level spectra that would reflect an internal electric field. This discrepancy implies that  besides the electronic reconstruction, extrinsic effects play a role, e.g. oxygen defects~\cite{Zhong:2010,Bristowe:2011}, adsorbates such as  water or hydrogen~\cite{Son:2010} or cation disorder~\cite{Willmott:2007,Qiao:2010} (for detailed reviews on the experimental and theoretical work see \cite{Huijben:2009,Pentcheva:PickettRev2010,Chen:Beigi:2010,Triscone:2011}).

The LAO/STO system  is not only of fundamental scientific interest, but is also a promising candidate for the development of electronics and spintronics devices~\cite{Mannhart:2010,cheng_NatNano2011}. For its incorporation in such devices the influence of metallic leads needs to be considered~\cite{Jany:2010,Bhalla:2010,Liu:cm}. Metallic overlayers have been investigated on a variety of perovskite surfaces, such as   \sto(001)~\cite{Asthagiri:Sholl:2002,mrovec:2009}, LaAlO$_3$(001)~\cite{Asthagiri:Sholl:2006,dong:2006} or BaTiO$_3$(001)~\cite{Sai:Rappe:2005,Stengel:Vanderbilt:Spaldin:2009}. Furthermore, the magnetoelectric coupling between ferromagnetic Fe and Co films and ferroelectric BaTiO$_3$ and PbTiO$_3$(001)-surfaces~\cite{Duan:Tsymbal:2006,Fechner:2008} has been explored. However, the impact of a metallic overlayer on a buried oxide interface has not been addressed so far theoretically. Based on DFT  calculations we show here that a metallic Ti overlayer not only acts as a Schottky barrier, but has a crucial influence on the electronic properties as it removes the internal electric field in the LAO film on STO(001).  Despite the lack of a potential build up, the underlying STO layer is metallic and a significantly enhanced carrier concentration emerges at the $n$-type interface as compared to the system without metallic contact. Although bulk Ti is nonmagnetic, the undercoordinated Ti in the contact layer shows an enhanced tendency towards magnetism with a significant spin polarization and a magnetic moment of 0.60~\muB. Most interestingly, quantum confinement within the STO-substrate can induce spin-polarized carriers at the interface. 

DFT calculations on $n$Ti/$m$LAO/STO(001), where $n$ ($m$) denotes the number of metallic overlayers (monolayers of LAO) were performed using the all-electron full-potential linearized augmented plane wave (FP-LAPW) method in the WIEN2k implementation \cite{Wien2k} and the generalized gradient approximation (GGA)\cite{GGA} of the exchange-correlation potential. Including  an on-site Coulomb correction within the LDA/GGA+$U$ approach \cite{Anisimov:Solovyev:etal1993} with $U=5$ eV and $J=1$eV applied on the Ti $3d$-states and $U=7$ eV on the La $4f$ states showed only small differences in electronic behavior.  The calculations are carried out using a symmetric slab with LAO and Ti layers on both sides of the STO-substrate and a vacuum region between the periodic images of at least $10$\AA. The lateral lattice parameter is set to the GGA equilibrium lattice constant of STO (3.92 \AA, slightly larger than the experimental value 3.905 \AA) and  the atomic positions are fully relaxed within tetragonal symmetry.

\begin{figure}[t!]
% \vspace{5mm}
   \begin{center}
\includegraphics[scale=0.43]{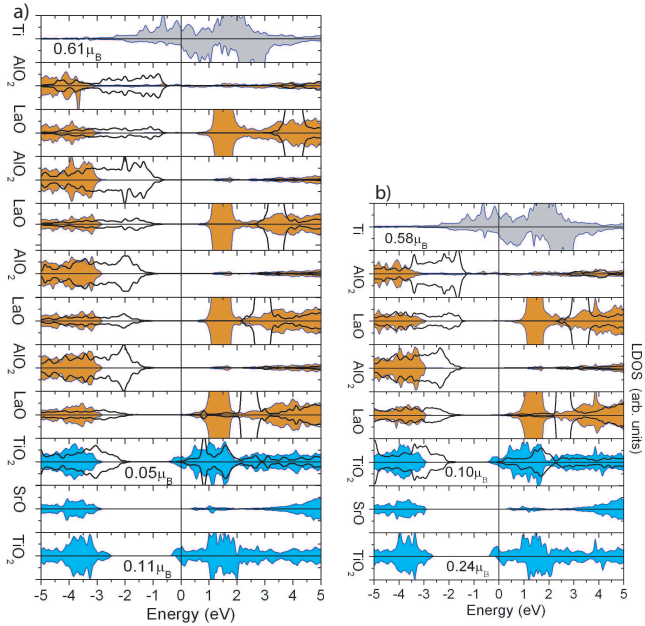}
   \end{center}
\caption{(Color online) Layer resolved density of states (LDOS) of a) 1Ti/4LAO/STO(001) and b) 1Ti/2LAO/STO(001). The potential build up in the uncovered systems (black line) is canceled in the ones with a Ti overlayer (filled area). Additionally, there is a significant spin-polarization both in the surface Ti, as well as the interface TiO$_2$ layers in 1Ti/$N$LAO/STO(001).  }\label{fig:dos1Ti2-4LAO3STO} 
\end{figure}
Fig.~\ref{fig:dos1Ti2-4LAO3STO}a shows the layer resolved density of states (LDOS) of a 1Ti/4LAO/STO(001) system, where the Ti atoms are adsorbed on top of the oxygen ions in the surface AlO$_2$ layer. A striking feature is that the electric field of the uncovered  LAO film (black line), expressed through an upward shift of the O $2p$ bands~\cite{Pentcheva:Pickett2009}, is completely eliminated  after the adsorption of the Ti overlayer. As a consequence of the vanishing electric field within the LAO layer, no dependence of the electronic properties  on the LAO thickness is expected, as opposed to the insulator-metal transition that occurs in the uncovered LAO/STO(001) system. Indeed the LDOS of 1Ti/2LAO/STO(001) (Fig.~\ref{fig:dos1Ti2-4LAO3STO}b, only 2 LAO layers) confirms a very similar behavior with no shifts of the O $2p$ bands  in the LAO part. %except for the topmost AlO$_2$ layer. 
Despite the lack of a potential build up, there is a considerable occupation of the Ti $3d$ band at the interface and thus both the surface Ti layer and the interface are metallic.  The charge transfer from the Ti adlayer  allows the Ti+LAO+STO system to equilibrate in charge and potential, with the result being that the Fermi level lies just within the STO conduction band.  

The low coordination of the surface Ti atoms enhances their tendency towards magnetism  resulting in a magnetic moment of 0.58\muB\ in  the surface layer. The electron gas at the interface is also spin-polarized with magnetic moments sensitive to the thickness of the LAO-spacer: for $m_{LAO}=4$ the magnetic moments are smaller (0.05/0.11\muB\  in the interface (IF)/IF-1 layer) than for  $m_{LAO}=2$ (0.10/0.24\muB\ in IF/IF-1).  Calculations  performed within GGA+$U$ for the latter case show a similar behavior  but with an even stronger  spin-polarization of  carriers and magnetic moments of Ti of 0.20/0.30 \muB\ in the IF/IF-1 layer, respectively. Thus correlation corrections have only small influence on the overall band alignment, confirming that the observed behavior is not affected by the well known underestimation of band gaps of LDA/GGA.

\begin{figure}[t]
%\vspace{2cm}
%\hspace{-2cm}
  \begin{center}
   \includegraphics[scale=0.5,angle=0]{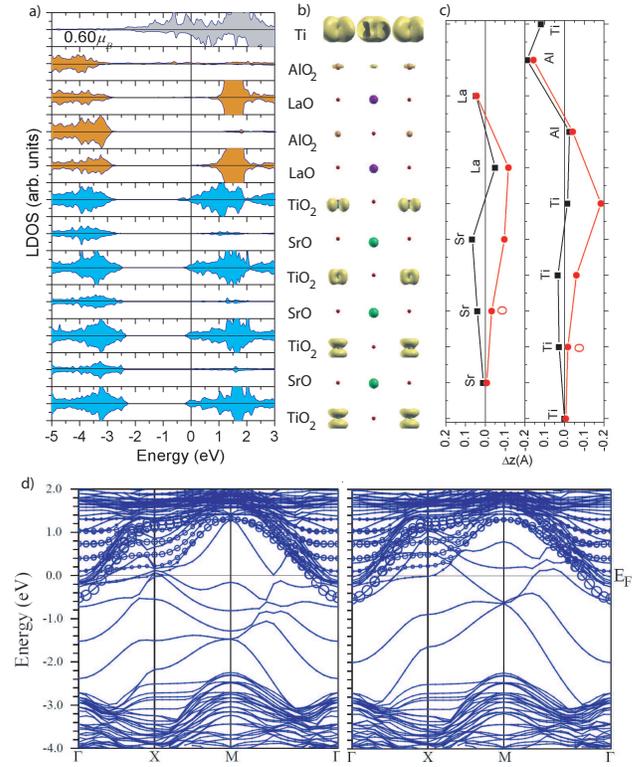}
   \end{center}
\caption{(Color online) a) Layer resolved density of states (LDOS) of 1Ti/2LAO/STO(001) within GGA with a 6.5ML thick STO-substrate; b) Side view of the system with the electron density integrated in the interval E$_{\rm F}-0.5$eV to E$_{\rm F}$; c) vertical displacements of cations and anions in 1Ti/2LAO/STO(001); d) majority and minority band structure where the $3d$ bands in the interface layer are emphasized with the $d_{xy}$ orbital being the lowest lying level at $\Gamma$. }\label{fig:1Ti2LAO6STO} 
\end{figure}

The calculations so far were performed with a rather thin substrate layer of 2.5 ML STO. To examine the dependence on the thickness of the substrate layer, we studied 1Ti/2LAO/STO(001) containing  a 6.5 ML thick STO part. As shown in Fig.~\ref{fig:1Ti2LAO6STO}a, the most prominent difference to the system with a thin STO layer is the suppression of spin-polarization of carriers at the interface, indicating that the spin-polarization is a result of confinement effects in the thin STO layer. Apart from this, a notable band bending occurs in the STO part of the heterostructure. The largest occupation of the Ti $3d$ band arises at the interface, followed by a decreasing occupation in deeper layers.  The electron density, integrated over states between E$_{\rm F}-0.5$eV to E$_{\rm F}$ in Fig.~\ref{fig:1Ti2LAO6STO}b, reveals orbital polarization of the Ti electrons in the conduction band: predominantly $d_{xy}$ character in the interface layer, nearly degenerate $t_{2g}$ occupation in IF-1, and a preferential occupation of $d_{xz}$, $d_{yz}$ levels in the deeper layers.   The band structure plotted in Fig.~\ref{fig:1Ti2LAO6STO}d shows that the conduction band minimum is at the $\Gamma$-point, formed by $d_{xy}$ states of Ti in the interface layer. While the  $d_{xy}$ bands  have a strong dispersion, the $d_{xz}$, $d_{yz}$ bands lie slightly higher in energy but are much heavier along the $\Gamma-X$ direction. In addition to the orbital polarization of the filled bands, the different band masses indicate a significant disparity in mobilities of electrons in the different $t_{2g}$ orbitals. Similar multiple subband structure has been recently reported  for  LAO/STO superlattices~\cite{Popovic:Satpathy:Martin:2008,Janicka:2009}, $\delta$-doped LAO in STO~\cite{Ong:2011} as well as doped STO(001)-surfaces~\cite{Santander:11,Meevasana:2011}. The remaining bands  between -2.5eV and E$_{\rm F}$ are associated with the surface Ti-layer.

%The altered band alignment across the LAO/STO interface affects also the structural relaxations: In the uncapped 
The altered boundary conditions in 1Ti/LAO/STO(001)  affect also the structural relaxations: In the uncapped LAO/STO(001) system a strong lattice polarization emerges  primarily in the LAO film  (with a buckling of $\sim0.26$\AA\  in the LaO and $\sim0.17$\AA\ in the subsurface AlO$_2$ layers) that   opposes the internal electric field~\cite{Pentcheva:Pickett2009}. As seen in the LDOS plots, the formation of a chemical bond between the metallic layer and the LAO film cancels the electric field within LAO, and hence, as shown in Fig.~\ref{fig:1Ti2LAO6STO}c,  cations and anions within the LAO film relax by similar amount with no appreciable buckling of the layers. On the other hand, the excess charge at the LAO/STO interface induces polarization in the STO substrate. The displacement between anions and cations is driven mainly by an outward oxygen shift and is largest at the interface ($0.17$\AA\ in the interface TiO$_2$ layer and  $0.16$\AA\ in the next SrO
  layer) and decays in deeper layers  away from the interface. This relaxation pattern resembles the one of  $n$-type LAO/STO and LTO/STO superlattices~\cite{Pentcheva:Pickett2008,Spaldin:2006}.
 
\begin{figure}[t]
 \hspace{-2cm}
   \begin{center}
   \includegraphics[scale=0.65,angle=0]{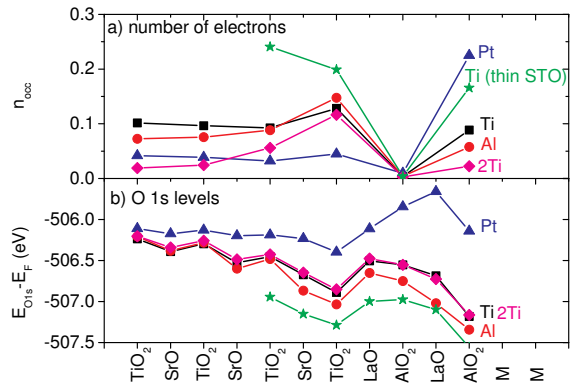}
   \end{center}
\caption{ a) Layer resolved electron occupation integrated between $E_F-0.65$ eV and $E_F$ for 1ML Ti and thin (green stars)/thick STO substrate (black squares); 2ML Ti (magenta diamonds), as well as 1ML Pt (blue triangles) and 1ML Al (red circles). Occupation within SrTiO$_3$ layers is due to Ti $3d$ electrons, the one in the topmost AlO$_2$ arises from metal induced gap states (MIGS).  Note that charge on the AlO$_2$ layer next to the interface is nearly vanishing in all cases, charge on the AO layers is zero (not shown). b) Positions of O$1s$ states with respect to $E_F$.} \label{fig:NoccO1s} 
\end{figure}

Increasing the thickness of Ti to 2 ML (with Ti in the second layer positioned above Al in the surface AlO$_2$-layer and La in the subsurface LaO-layer) leads to a significant reduction of the spin-polarization of the Ti film: the magnetic moment is 0.25\muB\ in the surface and -0.10\muB\ in the subsurface layer. This confirms that the high spin-polarization of Ti in 1Ti/$m$LAO/STO(001) is a result of the reduced coordination (for comparison, the magnetic moment of a free standing Ti layer is 0.90\muB). The weaker binding to the oxide surface in 2Ti/2LAO/STO(001) is expressed in a longer Ti-O bond length of 2.06 \AA\ compared to 2.00 \AA\ in 1Ti/2LAO/STO(001). Despite these differences in the structural and magnetic properties of the metallic overlayer, the occupation of the Ti $3d$ band at the interface is very similar (cf. Fig.\ref{fig:NoccO1s}a), but decreases quicker in the deeper layers.

  Besides Ti, we have studied also Pt and Al overlayers and find that the overall behavior -- strong reduction/cancellation of the electric field within LAO and metallization of the underlying STO -- is robust for all studied metallic contacts.   The layer resolved occupation of the Ti $3d$ band in \sto(001) (Fig.\ref{fig:NoccO1s}a) correlates with the positions of O$1s$ core levels with respect to the Fermi level (cf. Fig.\ref{fig:NoccO1s}b). The  lowest  O$1s$ eigenvalue occurs at the interface, with the strongst binding energy  for a Ti monolayer with a thin STO substrate (occupation of $0.2e^-$ of the Ti $3d$ orbitals). For the thicker STO substrate, the ordering of $3d$ occupation is Al, Ti, 2Ti with relatively small differences. Finally,  the lowest Ti $3d$ band  occupation is for a Pt contact with the respective O$1s$ eigenvalue at the interface being highest.  These quantitative differences are associated with the different chemical bond between the metal overlayer and the surface AlO$_2$ layer: \eg\ the bonding is strongly ionic in the case of Al with a significant charge transfer from Al to O  and is much weaker for a Pt overlayer with its nearly filled $4d$ states. The bond strength variation is also reflected in a sizeable difference in bond distances ($d_{\rm Al-O}=1.97$\AA, $d_{\rm Pt-O}=2.31$\AA). Furthermore, in the systems with a thick STO-substrate the O1s levels in deeper layers away from the IF shift upwards and converge to a similar value indicating that the inner potential relaxes to the bulk STO value. The total $3d$ band occupation is similar for Ti (thin and thick STO slab) and Al ($\sim0.4e$) and lower for 2Ti ($0.22e$) and Pt ($0.16e$).  %Such a relaxation does not occur for the system with thin STO substrate which also exhibits  a high occupation of the Ti $3d$ band in IF-1.  
\begin{figure}[t]
 \hspace{-2cm}
   \begin{center}
   \includegraphics[scale=0.5,angle=0]{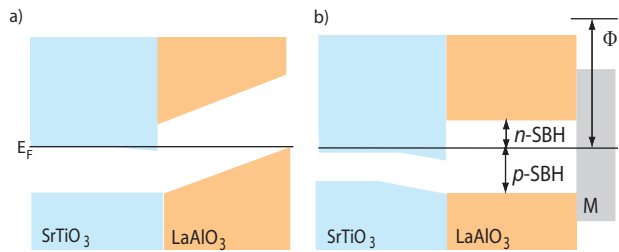}
   \end{center}
\caption{Schematic band diagram of a) LAO/STO(001) at the verge of an electronic reconstruction at the critical LAO thickness; b) LAO/STO(001) covered by a metallic  contact layer (M). Note, that the potential build up that leads to an electronic reconstruction in the uncovered LAO/STO(001) system at the critical thickness is  strongly reduced/eliminated in M/LAO/STO(001).}\label{fig:banddiagr} 
\end{figure}

The distinct mechanisms of formation of a q2DEG in  LAO/STO(001) with and without a  metallic contact are displayed in the schematic band diagram in  Fig.~\ref{fig:banddiagr}. For LAO/STO(001) a thickness dependent MIT occurs as a result of the potential buildup, where the electronic reconstruction involves the formation of holes at the surface and electrons at the interface.  In contrast, for $M$/LAO/STO(001) the potential in LAO is flat regardless of the LAO or STO thickness.  Simultaneously,  a q2DEG with higher carrier density is formed at the interface. Only in the case of Pt (and possiblly other noble metals), likely due to the weaker bonding and smaller charge transfer to the oxide layer, a small residual slope within LAO is found, consistent with the recently measured potential build up in Pt/LAO/STO(001)~\cite{Bhalla:2010}. 

The Schottky barrier height (SBH) between the metal and the oxide film is an important quantity which depends critically on the type of metal, the chemical bonding characteristics and the work function (see \eg\ ~\cite{mrovec:2009}): the $p$-SBH determined from the LDOS are  3.0 eV (Al), 2.8 eV (Ti) and 2.2 (Pt). The latter two values are slightly higher but show the same trend as recent results for Pt and Al on LAO(001)~\cite{dong:2006}. The corresponding $n$-SBH are  0.7 eV (Al), 0.9 eV (Ti) and 2.0 eV (Pt), determined from LDOS, or 1.6 eV higher in each case if the experimental band gap of LAO is considered. The conduction band offset ($n$-type SBH) is related to the effective work function of the system ($\Phi^{\rm Al}=3.53$ eV, $\Phi^{\rm Ti}=4.05$ eV,  and $\Phi^{\rm Pt}=5.60$ eV) and correlates also  with the  Ti $3d$ band occupation at the LAO/STO interface which is highest for an Al contact, followed by Ti and lowest for Pt.

These results show that metallic contacts ultimately change the electrostatic boundary conditions and represent a further powerful means besides oxygen defects\cite{Bristowe:2011} or adsorbates\cite{Son:2010}, oxide overlayers~\cite{Pentcheva:2010} or an external electric field~\cite{Cen:Thiel:Hammerl:etal2008,Bi:Levy:2010,Chen:2010} to tune the functionality  at the LAO/STO(001) interface. Despite analogies to the adsorption of hydrogen on LAO/STO(001)\cite{Son:2010}, there are also subtle differences (\eg\ the   strong dependence of the potential slope on coverage in the latter system). These differences emphasize  not only the importance of the  electrostatic boundary conditions\cite{Stengel:2011} but also of a precise understanding of structural effects and chemical bonding to LAO/STO(001) in order to achieve better understanding and control device performance.
\begin{acknowledgments}
We acknowledge discussions with J. Mannhart and financial support through the DFG SFB/TR80 (project C3)
and  grant {\sl h0721} for computational time at the Leibniz Rechenzentrum. V. G. R. L. acknowledges financial support from CONACYT (Mexico) and DAAD (Germany). W. E. P. was supported by U.S. Department of Energy Grant No. DE-FG02-04ER46111.
\end{acknowledgments}

%\bibliography{TiLAOSTOnv2}

\end{document}